\documentstyle[12pt]{article}  
  
\newcommand{\be}{\begin{equation}} % \label{#1#2#3}}     % non-hyper
\newcommand{\ee}{ \end{equation}}
\newcommand{\ba}{\begin{array}}
\newcommand{\ea}{\end{array}}
\newcommand{\bea}{\begin{eqnarray}}
\newcommand{\eea}{\end{eqnarray}}

\newcommand{\ft}[2]{{\textstyle\frac{#1}{#2}}}

\def\beq{\begin{equation}}
\def\eeq{\end{equation}}
\def\beqa{\begin{eqnarray}}
\def\eeqa{\end{eqnarray}}

\newcommand{\eqn}[1]{(\ref{#1})}

\newcommand{\pa}{\partial}

\newcommand{\bt}{\begin{tabular}}
\newcommand{\et}{\end{tabular}}

\begin{document}  
  
\begin{titlepage}  
%\flushleft{\today }  
\begin{flushright}  
HUB-EP-99/59\\
THU-99/35\\[1mm]
{\tt hep-th/9912225}
\end{flushright}  
  
\vspace{1cm}  
  
\begin{center}  
  
{\Large \bf{BPS Amplitudes, Helicity Supertraces and Membranes in M-Theory}}
  
\vspace{1.8cm}  

Bernard de Wit$^{1,2\,}$\footnote{bdewit@phys.uu.nl} and 
Dieter L\"{u}st$^{1,3\,}$\footnote{luest@physik.hu-berlin.de}
  
\vspace{.6cm}  
$^1${\em Institute for Theoretical Physics, University of California,
 \\ Santa Barbara, CA 93106, USA}

\vspace{.3cm}  
$^2${\em Institute for Theoretical Physics, Utrecht University,\\
Princetonplein 5, 3508 TA Utrecht, The Netherlands} 

\vspace{.3cm}  
$^3${\em Institut f\"{u}r Physik, Humboldt Universit\"{a}t zu Berlin,
\\ Invalidenstrasse, D-10115   Berlin, Germany}   

\vspace{.6cm}  
  
\begin{abstract}
We study BPS dominated loop amplitudes in M-theory
on $T^2$. For this purpose we generalize the concept of helicity
supertraces to nine spacetime dimensions. These traces distinguish
between various massive supermultiplets and appear as coefficients in
their one-loop contributions to $n$-graviton scattering
amplitudes. This can be used to show that
only ultrashort BPS multiplets contribute to the $R^4$ term in the
effective action, which was first computed by Green, Gutperle and
Vanhove. There are two inequivalent ultrashort BPS multiplets which
describe the Kaluza-Klein states and the wrapped membranes that cover
the torus a number of times. From the perspective of the type-II 
strings they correspond to momentum and winding states and D0 or D1
branes.  
\end{abstract}
  
\end{center}  
%%%%%%%%%%%%%%%%%%%%%%%%%%%%%%%%%%%%%%%%%%%%%%%%%% 
\vspace{.5cm} 
 
\flushleft{December 1999}
\end{titlepage}  
  
%%%%%%%%%%%%%%%%%%%%%%%%%%%%%%%%%%%%%%%%%%%%%%%%%%%%%%%%%%  
%\section{Introduction} 
M-theory is believed to provide a unifying framework
of all known superstring theories. Its low-energy limit is
described by eleven-dimensional supergravity \cite{CJS}. 
The latter theory exhibits nonrenormalizable ultraviolet behavior
which will presumably be cured once one includes the additional
M-theory degrees of freedom.  
Upon compactification to lower dimensions, all the duality symmetries
of string theory, such as S-, T- and U-duality, should become manifest.
However the fundamental, microscopic formulation of M-theory is so far
unknown. 
Matrix theory \cite{matrix1,matrix2} is one attempt in this direction. 
Closely related to this is the idea that supermembranes \cite{superm}
constitute the fundamental degrees of freedom of M-theory. 
Supermembrane theory may not suffer from the incompleteness of perturbative
string theory. Unlike string theory, which has both a string tension as
well as a coupling constant, it has no conventional perturbative expansion
as its only parameter is the membrane tension $T_{\rm m}$.

In this paper we consider BPS-dominated amplitudes\footnote{%%%%%%%%
  BPS amplitudes in perturbative and nonperturbative string theory
  have been discussed in many places -- see e.g.
  \cite{dkl,fklz,hm,curio,hma,koun,lerche}.} %%%%%%%%%%%%%%%%%%%%%%%%%
in M-theory compactified on a two-dimensional torus $T^2$, or
equivalently type-IIA/B superstring theory compactified on $S^1$. BPS
states play an important role in the computation of loop amplitudes in
theories with extended supersymmetry. The masses of the BPS states are
often supposed to be exact functions of the moduli so that amplitudes
that receive their contributions exclusively from BPS states are
also exact, even beyond perturbation theory. 
%The study of BPS states therefore provides a powerful method to derive 
%various nonrenormalization theorems. 
Here we will focus on the one-loop four-graviton
scattering amplitude which provides the coefficient of the
 gravitational  
$R^4$ term in the corresponding low-energy effective action. 
Computing this one-loop amplitude in eleven-dimensional supergravity
compactified on $T^2$ to nine spacetime dimensions
\cite{green,green1,green2,green3,green4}, the nonvanishing 
contribution is associated with the sum over the tower of
doubly-charged Kalazu-Klein states which circulate in the
nine-dimensional loop.  

{From} a nine-dimensional perspective one can study the one-loop
contributions 
to the $R^4$ term coming from a variety of supermultiplets. 
In order to analyze what kind of massive states 
can contribute to the loop amplitude, we adopt the concept
of helicity supertraces \cite{fsg,bk,kiritsis} but now extended 
to nine dimensions. We will then show that the $R^4$ terms are
exclusively generated by the ultrashort BPS multiplets. As was
recently dicussed  \cite{ADLN} there are two inequivalent ultrashort
multiplets, 
one corresponding to the Kaluza-Klein states from eleven-dimensional
supergravity and the other one corresponding to the Kaluza-Klein
states from IIB supergravity. The latter
states can be interpreted as the wrapped membrane states of
M-theory. Summing over both types of states thus yields the exact
answer for M-theory, as other massive supermultiplets cannot
contribute. We 
then show that including the sum over the wrapped membrane states
(using the same subtraction method as used in \cite{green} for the
Kaluza-Klein states on $T^2$) yields the correct amplitude consistent
with T-duality.  
The $R^6$ terms receive contributions from
intermediate BPS multiplets but not from long (non-BPS) multiplets,
while $R^8$ and higher-order terms receive contributions from all
supermultiplets 

Let us now first generalize the concept of helicity (super)traces for
massive states to
nine spacetime dimensions where the group of rest-frame rotations is given by 
SO(8). Following \cite{bk} we define a generating function for the
helicity traces by
\be
Z_{\bf r}(y) = {\rm Tr}_{\bf r}\, \Big(g_{\rm SO(8)}\Big) = {\rm
Tr}_{\bf r}\, \Big(\exp[ i\,\vec \phi\cdot \vec H]\, \Big)\,, 
\ee
where the trace and the SO(8) group element $g_{\rm SO(8)}$ are defined
in a representation $\bf r$. 
Obviously the generating function depends only on the group conjugacy
classes parametrized by four 
angles $\phi_i$ associated with the torus of the Cartan
subalgebra. The $H_i$ denote the generators of the Cartan
subalgebra in the representation $\bf r$ and the variables $y_i$ are
defined by $y_i= \exp[i\,\phi_i/2]$. The generating functions
satisfies the following properties, 
\bea
Z_{{\bf r}\oplus {\bf r^\prime}}(y) &=& 
Z_{\bf r}(y) + Z_{\bf r^\prime}(y)\,,    \nonumber\\ 
Z_{{\bf r}\otimes {\bf r^\prime}}(y)&=& 
Z_{\bf r} (y) \,Z_{\bf r^\prime}(y) \,,  \nonumber \\
Z_{\bf r}(y)&=&  Z_{\bf r} (y^{-1}) \,,  \nonumber \\
Z_{\bf r}(y=1)&=&  {\rm dim}({\bf r})    \,,
\eea
which are easy to prove. All the above results can readily be 
generalized to a supertrace for which we will not introduce any new
notation. Note, however, that the supertrace for $y_i=1$ is equal to
the graded trace of the identity and will thus vanish for
supermultiplets. 

The helicity traces are now $n$-th rank symmetric tensors defined by
(no summation over repeated indices)
\be
B_n({\bf r}) = {1\over 2^n}\; y_{i_1} {\pa\over\pa y_{i_1} } \,\cdots
\, y_{i_n} {\pa\over\pa y_{i_n} }  \;Z_{\bf r}(y) \Big\vert_{y_i=1}\,.
\ee
Observe that the trace will vanish for odd values of $n$. 
In the reduction to four spacetime dimensions, these results coincide
with those of \cite{bk}. 

Let us present the generating function for a number of relevant SO(8)
representations,
\bea
Z_{{\bf 8}_v}(y)  &=& \sum_i \,\Big(y^2_i + y^{-2}_i\Big)\,, \nonumber
\\ 
%%%%%%%%%%%%%%%%%
Z_{{\bf 8}_s}(y)  &=& \ft1{24}\! \sum_{i\not=j\not=k\not=l}
\Big(y_i\,y_j\,y_k\,y_l +
6\, y_i\,y_j\,y^{-1}_k\,y^{-1}_l+ 
y^{-1}_i\,y^{-1}_j\,y^{-1}_k\,y^{-1}_l \Big )\,, \nonumber \\
%%%%%%%%%%%%%%
Z_{{\bf 8}_c}(y)  &=& \ft1{6} \!\sum_{i\not=j\not=k\not=l}
\Big(  y_i\,y_j\,y_k\,y^{-1}_l +
y_i\,y^{-1}_j\,y^{-1}_k\,y^{-1}_l  \Big ) \,, \nonumber \\
%%%%%%%%%%%%%%%%
Z_{{\bf 28}_0}(y)   &=& \ft12 \sum_{i \not=j}\,  \Big(y^2_i\, y^2_j + 2
\,y^2_i\,y^{-2}_j  + y^{-2}_i\,y^{-2}_j\Big) + 4 \,, \nonumber \\
%%%%%%%%%%%%%%%%
Z_{{\bf 35}_0}(y) &=& \sum_i \,\Big(y^4_i + y^{-4}_i\Big) 
+ \ft12 \sum_{i \not=j} \,\Big(y^2_i\, y^2_j + 2 
\,y^2_i\,y^{-2}_j  + y^{-2}_i\,y^{-2}_j\Big) + 3 \,, 
 \nonumber \\
%%%%%%%%%%%%%%%%
Z_{{\bf 35}_0^{\prime}}(y) &=&  \ft1{24}\! \sum_{i\not=j\not=k\not=l}
\Big(y_i^2\,y_j^2\,y_k^2\,y_l^2 +
6\, y_i^2\,y_j^2\,y^{-2}_k\,y^{-2}_l+ 
y^{-2}_i\,y^{-2}_j\,y^{-2}_k\,y^{-2}_l \Big ) \nonumber \\
&&+ \ft12 \sum_{i \not=j}\,  \Big(y^2_i\, y^2_j + 2
\,y^2_i\,y^{-2}_j  + y^{-2}_i\,y^{-2}_j\Big) + 3 \,, \nonumber \\
%%%%%%%%%%%%%%%
Z_{{\bf 35}^{\prime\prime}_0}(y) &=& \ft1{6}\! \sum_{i\not=j\not=k\not=l}
\Big(  y_i^2\,y_j^2\,y_k^2\,y^{-2}_l +
y_i^2\,y^{-2}_j\,y^{-2}_k\,y^{-2}_l  \Big )  \nonumber \\
&& + \ft12 \sum_{i \not=j}\,  \Big(y^2_i\, y^2_j + 2
\,y^2_i\,y^{-2}_j  + y^{-2}_i\,y^{-2}_j\Big) + 3 \,, \nonumber \\ 
%%%%%%%%%%%%%%%
Z_{{\bf 56}_v}(y) &=& \ft16\! \sum_{i\not=j\not=k }
\Big(y^2_i\, y^2_j\, y_k^2
+ 3 \,y^2_i\,y^2_j\, y^{-2}_k + 3 \,y^2_i \,y^{-2}_j\, y^{-2}_k
 + y^{-2}_i\,y^{-2}_j\,y^{-2}_k \Big) \,, 
\nonumber\\ 
&& + 3 \,
\sum_i\, \Big(y^2_i + y^{-2}_i\Big)\,,  \nonumber \\
%%%%%%%%%%%%%%%
Z_{{\bf 56}_s}(y) &=&  \ft1{6}\! \sum_{i\not=j\not=k\not=l}
\Big(3\,y_i^3\,y_j\,y_k\,y_l^{-1} +
y_i^3\,y_j^{-1}\,y^{-1}_k\,y^{-1}_l+  y_i\,y_j\,y_k\,y^{-3}_l+ 
3\,y_i\,y^{-1}_j\,y^{-1}_k\,y^{-3}_l \Big ) \nonumber \\
&&+ \ft1{8} \sum_{i\not=j\not=k\not=l}
\Big(y_i\,y_j\,y_k\,y_l +
6\, y_i\,y_j\,y^{-1}_k\,y^{-1}_l+ 
y^{-1}_i\,y^{-1}_j\,y^{-1}_k\,y^{-1}_l \Big )\,,  \nonumber \\
%%%%%%%%%%%%%%%%%%
Z_{{\bf 56}_c}(y) &=&  \ft1{6}\! \sum_{i\not=j\not=k\not=l}
\Big(y_i^3 \,y_j\,y_k\,y_l +
3\, y_i^3\,y_j\,y^{-1}_k\,y^{-1}_l +
3\, y_i\,y_j\,y^{-1}_k\,y^{-3}_l + 
y^{-1}_i\,y^{-1}_j\,y^{-1}_k\,y^{-3}_l \Big ) \nonumber \\ 
&& + \ft1{2} \,\sum_{i\not=j\not=k\not=l}
\Big(  y_i\,y_j\,y_k\,y^{-1}_l +
y_i\,y^{-1}_j\,y^{-1}_k\,y^{-1}_l  \Big ) \,.
\label{SO(8)functions}
\eea
Note that the three representations ${\bf 35}_0$, ${\bf 35}_0^\prime$
and ${\bf 35}_0^{\prime\prime}$ appear in the square of the ${\bf
8}_v$, ${\bf 8}_s$ and ${\bf 8}_c$ representations,
respectively. These results may for instance be obtained by making use
of the weight vectors for these representations, which we have
collected in the appendix.

Let us now review the various possible BPS multiplets associated
with the nine-dimensional $N=2$ supersymmetry algebra with Lorentz invariant
central charges \cite{ADLN}. 
There are three independent charges. Two of them
rotate under the action of the the SO(2) automorphism group, while the
other one is invariant. 
Due to this particular structure of the central charges there exist
three types of BPS multiplets (as always we can
combine multiplets into larger multiplets with higher spins by
assigning spin to the Clifford vacuum, but this does not affect their 
characterization as BPS states). 

First there exist two
inequivalent types of ultrashort multiplets of massive 1/2 BPS states,
which are annihilated by inequivalent subsets of 16 supercharges. Both
multiplets contain $2^8=128+128$ states. When the SO(2) invariant
central charge vanishes and the other one has a magnitude equal to the
rest mass of the multiplet, the multiplet decomposes as 
%$a=\pm 1$ and  $b=0$ leads to 
\bea
({\bf 8}_v+ {\bf 8}_s)\times({\bf
8}_v+{\bf 8}_c)&=&\lbrack{\bf 1}_0+{\bf 8}_v+ {\bf 28}_0+ {\bf 35}_0
+{\bf 56}_v\rbrack_{\rm boson}\nonumber \\ 
&&+\, \lbrack {\bf 8}_s+{\bf 8}_c+
{\bf 56}_s+{\bf 56}_c\rbrack_{\rm fermion}.\label{shortsa} 
\eea
This is the multiplet
that comprises the Kaluza-Klein 
states of IIA supergravity compactified on $S^1$, which are the
momentum states of the compactified IIA 
string. Therefore this particular multiplet is known as the KKA
multiplet. Also the D0-branes of the IIA superstring transform
according to this multiplet. The second ultrashort multiplet is the
KKB multiplet. Now only the noninvariant central charge is different
from zero and equal in magnitude to the rest mass. The multiplet 
%obtained for $a=0$ and $b=\pm1$ and 
decomposes according to 
\bea
({\bf 8}_v+{\bf 8}_c)\times({\bf 8}_v+{\bf 8}_c)
&=&\lbrack{\bf 1}_0+{\bf 1}_0+{\bf 28}_0+{\bf 28}_0+ {\bf 35}_0 + {\bf 
35}_0^{\prime\prime}\rbrack_{\rm boson}\nonumber\\ 
&&+\, \lbrack {\bf 8}_s+{\bf
8}_s+{\bf 56}_s  +{\bf 56}_s\rbrack_{\rm fermion}. 
\label{shortsb}
\eea
This supermultiplet comprises the Kaluza-Klein states of IIB
supergravity compactified on $S^1$. 
Clearly, the BPS states associated with a membrane wrapped around
$T^2$ in eleven dimensions will also constitute KKB multiplets. 

Intermediate multiplets exist of massive 1/4 BPS states
annihilated by 8 supercharges. This multiplet carries both types of
central charges. The smallest multiplet contains
$2^{12}= 2^{11} +  2^{11}$ states. They do appear in string
theory as mixed states that carry both winding and momentum and have a
nonzero oscillator number in order to satisfy the mass-shell
condition. Hence they carry masses of the 
order of the string scale. The smallest multiplet associated with the
lowest spins decomposes as 
\bea
({\bf 8}_v+ {\bf 8}_s)\times ({\bf
 8}_v+{\bf 8}_c)\times({\bf 8}_v+{\bf 8}_c)
\label{intermed}
\eea
(or its conjugate). 

Finally there are the long (non-BPS) multiplets where all the
supercharges act nontrivially. The smallest one comprises $2^{16}$  
states and decomposes into 
\bea
({\bf 8}_v+ {\bf 8}_s)\times ({\bf
 8}_v+{\bf 8}_s)\times({\bf 8}_v+{\bf 8}_c)\times({\bf 8}_v+{\bf
8}_c).\label{long}
\eea   

Before proceeding to the determination of the generating functions for
the four classes of supermultiplets, let us define these functions for
two $N=1$ supermultiplets consisting of $({\bf 8}_v+ {\bf 8}_s)$ and 
$({\bf 8}_v+{\bf 8}_c)$. Observe that these are precisely the BPS
multiplets that one obtains from compactifying the ten-dimensional
supersymmetric gauge theory on $S^1$. We denote the corresponding
generating functions by $Z_{[{\bf s}]}$ and $Z_{[{\bf c}]}$. They are
given by\footnote{%%%%%%%%%%%%%%%%%%%%%%%%%%%%%%%%%%%%%%%%%%%%%%%%
 BPS supermultiplets are necessarily complex in order to transform
 under the central charge. This requires to add an additional factor
 equal to two in the definition of the  generating functions and the
 helicity traces. We suppress this factor throughout because we
 eventually sum over the (positive and negative) charges}%%%%%%%%%  
\bea
Z_{[{\bf s}]}(y)  &=& Z_{{\bf 8}_v}(y)- Z_{{\bf 8}_s}(y)  \,,
\nonumber \\ 
Z_{[{\bf c}]}(y)  &=& Z_{{\bf 8}_v}(y) - Z_{{\bf 8}_c}(y) \,,
\label{N=1fct}
\eea
and can be calculated from the expressions given in
\eqn{SO(8)functions}. It then follows
straightforwardly that the helicity traces generated by $Z_{[{\bf
s}]}$ and $Z_{[{\bf c}]}$ vanish for $n<4$. 

{From} the above result it follows that 
\bea
Z_{\rm KKA} (y) &=& Z_{\bf r}(y)\, Z_{[{\bf s}]}(y) \, Z_{[{\bf c}]}(y)
\,,\nonumber \\ 
Z_{\rm KKB} (y) &=& Z_{\bf r}(y)\, Z^2 _{[{\bf c}]}(y) \,,\nonumber \\  
Z_{\rm INTERM} (y) &=& Z_{\bf r}(y)\, Z_{[{\bf s}]}(y) \, 
Z^2_{[{\bf c}]}(y)\,,\nonumber \\  
Z_{\rm LONG} (y) &=& Z_{\bf r}(y)\, Z^2_{[{\bf s}]}(y) \, 
Z^2_{[{\bf c}]}(y)\,, 
\eea
where $\bf r$ denotes the spin representation of the Clifford
vacuum. 
This suffices to show that the helicity supertraces $B_n$ are
vanishing for KKA and KKB multiplets whenever $n<8$, for intermediate
multiplets whenever $n<12$ and for long multiplets whenever $n<16$.
This result has important implications for the one-loop graviton
amplitudes in type-II string theories, as we will show below
(obviously there is a corresponding result for $N=1$ theories
pertaining to the one-loop gauge field amplitudes). 

Subsequently we review the spectrum of the BPS states of M-theory
compactified to nine spacetime dimensions on $T^2$ \cite{JHS1,RT,ADLN}. 
The BPS mass formula based on the supersymmetry algebra
with a membrane winding charge, takes the form 
\bea
M_{q_1,q_2,p} =  {1\over {\sqrt{A\, \tau_2}}}\, \vert q_1 -\tau\,q_2\vert + 
T_{\rm m}\,{A}\,\vert p\vert  \,. \label{membrane-bps}
\eea
The ultrashort BPS multiplets are either given by Kalazu-Klein states
with momentum numbers $(q_1,q_2)$ along the two periods of $T^2$ or by
wrapped membranes which cover the torus $p$ times. 
The Kaluza-Klein charges $(q_1,q_2)$ transform as doublets under the
$SO(2)$ automorphism group of the supersymmetry algebra; the charge
associated with the wrapped membrane states is invariant under this
group. In \eqn{membrane-bps} $A$ denotes the volume of $T^2$,
%$A=\sqrt{\det G_{ij}}$, $G_{ij}$ being its metric 
measured with respect to the 11-dimensional metric, $\tau= \tau_1 +
i\tau_2$ denotes the complex structure of $T^2$ and $T_{\rm m}$
denotes the membrane tension. 

The BPS mass formula \eqn{membrane-bps} can be interpreted in the
context of type-II superstrings compactified on $S^1$. 
For this purpose let us recall the well-known relations
between the string and M-theory parameters \cite{witten,JHS}.
For simplicity we set $T_{\rm m}=1$ and assume that the M-theory
metric is diagonal and has 
the form $G_{ij}={\rm diag}\,(r_9^2,r_{10}^2)$, so that $A=r_9r_{10}$,
$\tau_1=0$ and $\tau_2=r_9/ r_{10}$. The ten-dimensional IIA
string coupling and the nine-dimensional radius of type-IIA on $S^1$
(in the string frame) are related to the M-theory parameters by 
\bea
g_{10}^A=r_{10}^{3/2},\qquad
R_9^A=r_9\,\sqrt{r_{10}}\,.\label{rnine}
\eea
Using T-duality between IIA and IIB in nine dimensions we can express
the corresponding IIB parameters in the following way (everything
measured with respect to the string metric in units of the string scale
$1/\sqrt{\alpha '}$),
\bea
R_9^B=\frac{1}{R_9^A}=\frac{1}{r_9\, \sqrt{r_{10}}},\qquad
g_{10}^B=\frac{1}{\tau_2}=\frac{r_{10}}{r_9}.\label{rb}
\eea
Subsequently we can express the BPS mass formula \eqn{membrane-bps} in
the string frame  
in terms of IIA and IIB string theory variables, respectively, 
\bea
M_{q_1,q_2,p}
=\Big\vert \, {q_1\over R_9^A} +{q_2\over g_{10}^A}\, \Big\vert 
+\vert \, p \, \vert R_9^A
=
\Big \vert \, q_1R_9^B+q_2{ R_9^B\over g_{10}^B}\,\Big \vert 
+{\vert \, p\, \vert\over R_9^B}
\, .
\label{ma}
\eea
Therefore from the perspective of the IIA string theory, $q_1$ is
the IIA Kaluza-Klein momentum number, while $q_2$ is the D0-brane charge.
The M-theory membrane wrapping number $p$ becomes the perturbative
winding number in the IIA string. On the other hand, 
from the IIB perspective, $q_1$ and $q_2$ 
are the winding numbers of the elemenary string  and of the solitonic
D1 string, and the membrane wrapping number $p$ is the IIB
Kaluza-Klein momentum. The IIB strong-weak coupling S-duality
interchanges the elementary strings with the D1 strings:
$q_1\leftrightarrow q_2$. Under the IIA/B duality,
$q_1\leftrightarrow p$, the IIA/IIB mass formulas are interchanged
provided one interchanges the D0 with the D1 states. 
Clearly the presence of the wrapped membranes is essential, 
as they correspond either to the IIA winding states or to the IIB
momentum states, respectively.  

In (perturbative) type-II string theory the various BPS states emerge
as follows. Ultrashort BPS states,
which preserve 16 of the total 32 supersymmetries, must be 1/2 BPS
states both with respect to the left-moving $N=1$ and also with respect to
the right-moving $N=1$ supersymmetry algebra. Therefore these states
are not allowed to carry any oscillator excitations, i.e.
$N_L=N_R=0$. Hence their SO(8) helicities are entirely
determined by the bosonic and fermionic groundstates, $({\bf 8}_v+
{\bf 8}_{s(c)})_L\times ({\bf 8}_v+{\bf 8}_c)_R$, leading to the 
decomposition in \eqn{shortsa} and \eqn{shortsb} for the two
ultrashort multiplets. Using the string level matching condition, 
$p_L^2=p_R^2$, where $p_{L,R}=q_1/ R_9^A \pm  pR_9^A$ for IIA, and
similarly for IIB, it follows that the ultrashort BPS states can carry
either  nonvanishing KKA quantum numbers $q_i$ or non-vanishing KKB
quantum numbers $p$, but not both. 

The intermediate, 1/4 BPS states are short 1/2 BPS multiplets with
respect to the left-moving $N=1$ supersymmetry algebra but are long
multiplets with respect to the right-moving $N=1$ supersymmetry
algebra (or vice versa). Therefore they require $N_L=0$ and $N_R$ 
arbitrary. The level matching condition $p_L^2=p_R^2+2 N_R$ now tells
us that intermediate multiplets must have both winding and momentum
and tus carry both type of charges. For $N_R=1$  we have
precisely the decomposition \eqn{intermed}, while for higher $N_R$ we
have intermediate BPS multiplets of higher spin. 
Finally the long multiplets have $N_L,N_R\neq 0$. The smallest 
long multiplet with $N_L=N_R=1$ has precisely the decomposition
(\ref{long}) and others will have higher spins. 

After this perusal of the various supermultiplets from complementary
viewpoints, we briefly discuss the helicity generating partition function of
M-theory on $T^2$. For string theory it has been argued
\cite{kiritsis} that this 
partition function is related to the string amplitudes for
multi-graviton scattering which we will turn to shortly. Hence we
start from the corresponding expression for the perturbative type-II 
string in nine dimensions,
\bea
Z(\phi,\bar\phi; q,\bar q)={\rm Str}\,\Big[ q^{L_0}\bar q^{\bar L_0}
e^{i\phi_iH_i+ i\bar\phi_i\bar H_i}\Big] \, .
\eea
Here $q$ and $\bar q$ are related to the world-sheet modular parameter
in the usual way.  
The expression for $Z(\phi,\bar\phi;q,\bar q)$ can be written as
\bea
&&Z(\phi,\bar\phi;q, \bar q) \sim
{1\over| \eta(q)|^{8}}\;\Gamma_{1,1}(q,\bar q)\; 
%\prod_{i=1}^4\,
%\vert \xi(\phi_i,q)\vert^2 
%\nonumber 
\\
 &&\hspace{6mm} \times 
 \Big[-\prod_{i=1}^4
\theta_1(\phi_i,q)  
+\prod_{i=1}^4\theta_2(\phi_i,q)-
\prod_{i=1}^4\theta_3(\phi_i,q)+
\prod_{i=1}^4\theta_4(\phi_i,q)\Big]  \nonumber \\
&& \hspace{6mm} \times \Big[{\rm sgn}(p_L\, p_R)\,
\prod_{i=1}^4\theta_1(\bar\phi_i,\bar q)
+\prod_{i=1}^4\theta_2(\bar \phi_i,\bar q)-
\prod_{i=1}^4\theta_3(\bar \phi_i,\bar q)+
\prod_{i=1}^4\theta_4(\bar \phi_i,\bar q)\Big]  \,, \nonumber
\label{zstring}
\eea
where 
%$\xi(\phi_i,q)= (\sin \pi\phi_i/\pi) \, 
%(\theta_1^\prime(\phi_i,q)/\theta_1(\phi_i,q))$ 
%counts the helictity contribution of the bosonic oscillators (where
%$\theta^\prime_1 = \pa \theta_1 /\pa \phi$); 
\bea
\Gamma_{1,1}(q,\bar q)=\sum_{p_L,p_R}\; q^{p_L^2/2} \;
\bar q^{p_R^2/2}  \, . \label{lattice}
\eea
Note that in $Z(\phi,\bar\phi; q,\bar q)$
we have neglected the contributions from the bosonic oscillators 
(so that we will not discuss the issue of the modular transformation
properties of $Z$). 
The products of $\theta$-functions originate from the sum over spin
structures and encode the helicities associated with the fermionic
zero modes;
%The result for \eqn{zstring} follows from standard calculations
%(see e.g. \cite{kiritsis}) but there are some features which are
%new. 
in particular we draw attention to the presence of the sign 
factor ${\rm sgn}(p_L\,p_R)={\rm sgn}(q_1^2/(R_9^A)^2-p^2(R_9^A)^2)$ 
in front of the
$\theta_1(\bar\phi_i,\bar q)$ functions, which is required in order to
correctly account for the different helicities of the momentum and
winding states \cite{ADLN}. Obviously the lattice summation in
\eqn{lattice} acts on this sign factor as well.
Owing to the sign factor, which is $+1$ for IIA momentum states and 
IIB winding states, respectively, and
$-1$ for IIA winding states and IIB momentum states, respectively, the
partition function \eqn{zstring} tends to the correct expression for
IIA string theory in the decompactification
limit $R_9^A\rightarrow\infty$ and for IIB string theory in the limit
$R_9^A\rightarrow 0$. 

An intriguing question is whether one can extend the sum over the
perturbative momentum and winding string states, by including the
nonperturbative BPS states of M-theory with masses as given in
\eqn{membrane-bps}. 
By using the SO(2) symmetry among the Kaluza-Klein states 
with momentum numbers $q_1$ and $q_2$, we are led to 
postulate that the same split into left- and right-moving masses $L_0$
and $\bar L_0$ as in string theory will also hold for compactified
M-theory, i.e., we replace $p_L$ and $p_R$ in the above expression by 
\be
{p_{L,R}\over \sqrt2}  = {q_1-\tau
q_2\over\sqrt{A\,\tau_2}}\pm AT_{\rm m}\,p \;. 
\ee
Then the M-theory BPS sum takes the following form:
\bea
\Gamma_{1,1}(q,\bar q)=\sum_{q_1,q_2,p}\; q^{{\textstyle\vert}{q_1-\tau
q_2\over\sqrt{A\tau_2}}+AT_{\rm m}\,p{\textstyle\vert}^2} \;
\bar q^{{\textstyle \vert} {q_1-\tau q_2\over\sqrt {A\tau_2}}-AT_{\rm m}\,p
{\textstyle\vert}^2}\, . 
\eea

However, $\Gamma_{1,1}$ does not depend on the angles and therefore
plays only a minor role on what follows. Clearly, apart from
$\Gamma_{1,1}$,  the expression \eqn{zstring} factorizes in a
holomorphic and an anti-holomorphic part. 
The appearance of the sum over the powers of theta-functions,
$\prod_{i=1}^4\theta_a(\phi_i)$, is of crucial importance for what
follows. Using  the Riemann identity the $\theta$-functions, we can
rewrite  (\ref{zstring}) in the following way:
\bea
Z(\phi,\bar\phi;q,\bar q)=  \,{1\over |\eta(q)|^8}
\, \Gamma_{1,1}(q,\bar q)\; \prod_{i=1}^4\,  \theta_1(\phi_i'/2,q)\,
\theta_1(\bar\phi_i'/2,\bar q) \, ,
\label{zstringa}
\eea
where 
\bea
\begin{array}{rcl}
\phi_1'&=&-\phi_1+\phi_2+\phi_3+\phi_4\,,\\
\phi_2'&=&-\phi_1+\phi_2-\phi_3-\phi_4\,,\\
\phi_3'&=&-\phi_1-\phi_2+\phi_3-\phi_4\,,\\
\phi_4'&=&-\phi_1-\phi_2-\phi_3+\phi_4\,,
\end{array}
\qquad
\begin{array}{rcl}
\bar \phi_1'&=&{\rm sgn}(p_L\,p_R)\,\bar\phi_1+
\bar\phi_2+\bar\phi_3+\bar\phi_4\,,\\
\bar \phi_2'&=&{\rm sgn}(p_L\,p_R)\,\bar\phi_1+
\bar\phi_2-\bar\phi_3-\bar\phi_4\,,\\
\bar \phi_3'&=&{\rm sgn}(p_L\,p_R)\,\bar\phi_1-
\bar\phi_2+\bar\phi_3-\bar\phi_4\,,\\
\bar \phi_4'&=&{\rm sgn}(p_L\,p_R)\,\bar\phi_1-
\bar\phi_2-\bar\phi_3+\bar\phi_4\,. 
\end{array}
\eea

The holomorphic part of the helicity generating partition function is
the direct string generalization of the field theory generating function  
$Z_{{\bf 8}_v}(y)-Z_{{\bf 8}_{c}}(y)$,
discussed before, where in the string case
we deal not only with the
graded sum over the vector ${\bf 8}_v$ and the spinor
${\bf 8}_{c}$ ground states, but also with the graded sum
over the whole weight lattice related to these two $SO(8)$
representations. This sum vanishes for $\phi_i=0$ due
to space-time supersymmetry.
The corresponding helicity supertraces are given by the multiple
derivatives of the partition function, after setting
$\phi_i=\bar\phi_i$, i.e., 
\bea
B_{2n}(q)=(-1)^n\,{\partial\over\partial\phi_{i_1}}\cdots{\partial\over
\partial\phi_{i_{2n}}}\,
%{\partial\over\partial\bar\phi_{i_1}}\dots
%{\partial\over\partial\bar\phi_{i_n}}
Z(\phi_i, \phi_i; q,\bar q)\Big|_{\phi_i=0}\, .
\eea
Because $\theta_1(0,q)=0$ and
$\partial\theta_1(\phi,q)/\partial\phi
|_{\phi=0}=-2\pi\eta(q)^3$,  we
need at least four powers of $\partial /\partial\phi_i$ 
and four powers of $\partial /\partial\bar\phi_i$
in order to get a nonvanishing result, in agreement with
the previous discussion.
For example, $B_8$ can only get contributions from  ultrashort KKA or
KKB multiplets, while 
$B_{12}$ receives contributions from both ultrashort and intermediate
supermultiplets. 

Now we turn to the evaluation of the $n$-graviton
amplitudes, i.e. we consider one-loop string amplitudes with $n$
insertions of graviton vertex operators. In the  
zero-ghost picture, the graviton vertex operator has the form, 
\bea
G_{mn}\sim \Big(\partial X_m + ip_m\sum_{i=1}^4\psi_i\psi_{-i}\Big)
\Big(\bar\partial\bar X_n + ip_n\sum_{i=1}^4\bar\psi_i\bar\psi_{-i}\Big)
\,{\rm e}^{i\,p\cdot X}\,,
\eea
Hence the graviton amplitude can be expanded in terms of powers of the
operators $\psi_i\psi_{-i}$ and $\bar\psi_i\bar\psi_{-i}$ which are
the Cartan subalgebra currents of SO(8) in the left- and right-moving
sector generated by the world-sheet fermions. Their charges 
$H_i= (2\pi i)^{-1}\oint \psi_i\psi_{-i}$ and $\bar H_i= (2\pi i
)^{-1}\oint\bar\psi_i\bar\psi_{-i}$ are the Cartan algebra
generators of SO(8). Furthermore the amplitude factorizes into
contributions from the left- and the right-moving sector. Each
contribution thus decomposes into a linear combination of the helicity
supertraces that are generated by the functions \eqn{N=1fct}.
Because the number of gravitons is equal to the highest power of
of helicity operators $H_i$ and $\bar H_i$,
the $n$-graviton amplitude is proportional to $B_{2n}$ and lower
helicity supertraces. 
This proves that the four-graviton amplitude must be
proportional to the helicity supertrace $B_8$ and will thus receive 
constributions from only ultrashort BPS multiplets. Likewise the
six-graviton amplitude will receive contributions from both ultrashort
and intermediate BPS multiplets. The contribution from the latter
contains a term proportional to $B_{12}$. For eight and more gravitons all
supermultiplets will in principle contribute. 

Having established these results, we briefly consider the $R^4$
terms in more detail. Here 
%In principle, {\it one can perform this calculation in
%string theory}. We will comment on this later, but first 
we cast the
calculation in terms of nine-dimensional field theory, where the
relevant amplitude has the structure of a
box diagram in massive $\varphi^3$ theory in nine spacetime
dimensions. In view of our previous results it suffices to restrict
ourselves to contributions from the KKA 
and KKB states. Since we evaluate the coefficient of the $R^4$ term at
zero momentum, such an integral takes the form 
\be
{1\over (2\pi)^9}\int\;{d^9q \over (q^2+M^2)^4} = {1\over 6}\,{1\over
(4\pi)^{9/2} } \,  \int_0^\infty \,dt\;   
t^{-3/2} \;{\rm e}^{-t\,M^2} \,. \label{9d-integral}
\ee
Observe that this integral has a linear ultraviolet divergence, which
reflects itself in the singular behaviour of the integrand at $t=0$. 
The integral has the dimension of a mass, which is
appropriate for a coupling constant of $R^4$ in nine spacetime
dimensions. We will now use \eqn{9d-integral} and the BPS mass formula
\eqn{membrane-bps} and determine the contributions from both the KKA
and the KKB states.
For the former we obtain
\bea
A_4^{\rm KKA} &=& {1\over 6}\,{1\over
(4\pi)^{9/2} } \,  \int_0^\infty \,dt\;   
t^{-3/2} \;\sum_{q_1,q_2}\,{\rm e}^{-t\,(A\,\tau_2)^{-1} \vert
q_1-\tau q_2\vert^2} \nonumber \\
&=& {1\over 3}\,{A^{-1/2}\over
(4\pi)^{5} } \,  \int_0^\infty \,dt\;   
t^{1/2} \;\sum_{q_1^\prime,q_2^\prime}\,{\rm
e}^{-\pi\,t\,\tau_2^{-1} \vert 
q_1^\prime +\tau q_2^\prime\vert^2} \nonumber \\
&=& {2\over 3}\,{A^{-1/2} \over
(4\pi)^6 } \; f(\tau,\bar \tau) \,, \label{4KKA}
\eea
where the modular function $f(\tau,\bar\tau)$ is defined by \cite{green} 
\be 
f(\tau,\bar \tau) =  \sum_{(q_1^\prime,q_2^\prime)\neq
(0,0)}\,{\tau_2^{3/2}\over  \vert q_1^\prime+\tau q_2^\prime \vert^3}
\,.
\ee
In the second line of \eqn{4KKA} we performed a Poisson resummation and
changed the integration variable $t\to \pi A/t$. Observe that the
ultraviolet divergence is now associated with the upper integration
boundary and is only present for the term with $q_1^\prime=q_2^\prime=0$.
This contribution was dropped in the third line. 

Likewise, we obtain for the KKB states, 
\bea
A_4^{\rm KKB} &=& {1\over 6}\,{1\over
(4\pi)^{9/2} } \,  \int_0^\infty \,dt\;   
t^{-3/2} \;\sum_p\,{\rm e}^{-t\,T_{\rm m}^2 \,A^2\, p^2} \nonumber \\
&=& {1\over 3}\,{T_{\rm m}\, A \over
(4\pi)^5} \,  \int_0^\infty \,dt\;   
\;\sum_{p^\prime}\,{\rm e}^{-\pi\,t\,p^{\prime\,2}} \nonumber \\
&=& {4\over 3}\,{T_{\rm m}\, A  \over
(4\pi)^6 } \; \sum_{p^\prime\neq 0}\,{1\over  p^{\prime \,2}} 
\,. \label{4KKB}
\eea
Again we performed a Poisson resummation and changed the integration
variable  $t\to \pi/ t\,( T_{\rm m} A)^2$; and we dropped the
ultraviolet divergence associated with 
$p^\prime =0$ in the third line. The sum over $p^\prime$ in the third
line is equal to $2\zeta(2)= \pi^2/3$. We return to the issue of
ultraviolent divergences shortly. 

Combining the results \eqn{4KKA} and \eqn{4KKB} gives rise to
\be
A_4^{\rm KKA+KKB} = {2\over 3}\,{1 \over
(4\pi)^6 } \; \Big[ A^{-1/2} \, f(\tau,\bar \tau) + \ft23
\pi^2\,T_{\rm m}\,A \Big] \,. \label{4KKAB}
\ee
This result is invariant under the IIB S-duality symmetry 
$\tau\rightarrow (a\tau +b)/(c\tau+d)$. The contribution from the KKB
states is such that the result is compatible with T-duality of type-II
string theory.

Let us consider the two possible decompactification limits to the
ten-dimensional IIA/B theories. Here we need the result that, for
$\tau_1=0$ and $\tau_2$ large, the function $f(\tau,\bar\tau)$ has 
the form 
\bea
f(\tau,\bar\tau)=2\,\zeta(3)\,\tau_2^{3/2}+ \ft23\pi^2\,\tau_2^{-1/2}+
\cdots \,, \label{fexpansion}
\eea 
up to terms that are exponentially suppressed. We can now use the
relations between string and M-theory parameters and rewrite the
expression \eqn{4KKAB} as 
\be
A_4^{\rm KKA+KKB} = {4\over 3}\,{\sqrt{r_{10}}\,R_{9}^B \over
(4\pi)^6 } \; \left[  \left ( {\zeta(3) \over (g_{10}^B)^2}
+ \ft13 \pi^2+ \cdots \right)   + \ft13 \pi^2\,{1\over
(R_{9}^B)^2}\right] \, . \label{B-decomp}
\ee
%where the ellipses denote exponentially suppressed terms in
%$(g_{10}^B)^{-2}$, which correspond to nonperturbative IIB D-instanton
%contributions. 
For
large $R_{9}^B$ the last term vanishes while the first two terms yield
the tree and the one-loop contributions to the $R^4$ term, 
and the ellipses denote exponentially suppressed terms in
$(g_{10}^B)^{-2}$, which correspond to nonperturbative IIB D-instanton
contributions.
(The factor
$\sqrt{r_{10}}$ is related to the fact that we expressed the KKA and
KKB masses in the M-theory frame; in the string frame this factor
disappears.)   

%and the dots denote  
%exponentially suppressed terms in $(g_{10}^B)^{-2}$, which
%correspond to nonperturbative IIB D-instanton contributions
%be  is absorbed into the ten-dimensional metric determinant, 
%$\sqrt {G_{10}^B}=\sqrt{G_9^B}R_9^B$, 

In the IIA decompactification limit $R_9^A\rightarrow\infty$ we obtain 
\be
A_4^{\rm KKA+KKB} = {4\over 3}\,{\sqrt{r_{10}}\,R_{9}^A \over
(4\pi)^6 } \; \left[  \left ( {\zeta(3) \over (g_{10}^A)^2}
+ \ft13 \pi^2 {1\over (R_9^A)^2} + \cdots \right)   + \ft13 \pi^2
\right] \,. \label{A-decomp}
\ee
Now the ellipses denote terms that are  exponentially suppressed 
in $(R_9^A/g_{10}^A)^2$. Dropping these terms as well as the second term 
term which all vanish in the decompactification limit, we are left 
with the IIA string tree and one-loop contribution to $R^4$ (the 
latter originated from the
KKB multiplets). Hence perturbative T-duality is manifest is manifest
in the results 

It is an intriguing question why the subtraction method based on
Poisson resummation and the subsequent subtraction of the $q_1^\prime=
q_2^\prime=0$ and $p^\prime=0$ terms leads to the correct result. A
puzzling feature of the calculation is that the sum over the KKA and
the sum over the KKB states both include the massless states, which
seems to make no sense. Correcting for this requires to subtract an
infinite term which could in principle be cancelled against the
positive infinite contributions of the massive BPS states. However,
why the cut-off should be fine-tuned such that this cancellation takes
place remains a mystery. Within the context of nine-dimensional
supersymmetry there are no other states that could possibly cancel the
infinite terms.
The coefficient of the $R^4$ term is not generically
finite, which reflects itself in the fact that the KKA contributions
disappear in the decompactification limit to eleven dimensions,
$A\rightarrow\infty$, which is somewhat counterintuitive. 
We stress that supersymmetry does not seem to be relevant for obtaining
finite results, in view of the fact that the supertraces for
the KKA and KKB multiplets do not vanish, and no cancellation seems to arise 
between the inifinite contributions from the two types of
multiplets. 

%%%%%%%%%%%%%%%%%%%%%%%%%%%%%%%%%%%%%%%%%%%%%%%%%%%%%%%%%%%%%%%%%%%
\vskip 8mm
\noindent {\bf Acknowledgements} \\
We thank Mohab Abou-Zeid, Elias Kiritsis, Wolfgang Lerche, Hermann Nicolai
and Kostas Skenderis for clarifying discussions. 
This research was supported in part by the National Science Foundation
under Grant No. PHY94-07194 through the Institute for Theoretical
Physics in Santa Barbara. We thank the institute for the hospitality
extended to us during this work. The research was also supported by the
European Commission TMR Program under 
contract ERBFMRX-CT96-0045, in which Humboldt University at Berlin and
Utrecht University are associated.

\vskip 8mm
\noindent
{\bf Appendix}\\
%%%%%%%%%%%%%%%%%%%%%%
\noindent
For the convenience of the reader we present the weight vectors for a
number of SO(8) representations which can be used in the construction
of the generating functions \eqn{SO(8)functions}. To each of the
weight vectors one 
must add all possible permutations. In certain cases the number of
minus signs must be even or odd. This is indicated by, e.g. $[+ 20;{\rm
even}]$, implying  that 20 independent  permutations should be added
with an even number of minus signs. The subscripts $0$, $v$, $s$ and
$c$ refer to the different conjugacy classes of weight vectors. 
\beqa
{\bf 8}_v:&\quad&  \vec\lambda=(\pm 1,0,0,0)\quad [+6] \,;\nonumber\\
{\bf 8}_s: &\quad&  \vec\lambda= 
(\pm\ft12,\pm\ft12,\pm\ft12,\pm\ft12 ) \quad [{\rm even}]\, ;  \nonumber\\
{\bf 8}_c: &\quad&  \vec\lambda=(\pm\ft12,\pm\ft12,\pm\ft12,
\pm\ft12) \quad [{\rm odd}] \, ;\nonumber\\
{\bf 28}_0: &\quad&  \vec\lambda=(\pm 1,\pm 1,0,0) \quad [+ 20]\,, \quad 
  \vec\lambda_{1,2,3,4}^\prime=(0,0,0,0)\, ;\nonumber\\
{\bf 35}_0: &\quad&  \vec\lambda= (\pm 2,0,0,0)\quad [+ 6]\, ,\quad
\vec\lambda'=(\pm 1,\pm 1,0,0)\quad [+20] \,,\nonumber \\
&{}& \vec\lambda_{1,2,3}^{\prime\prime}=(0,0,0,0)\, ;\nonumber\\
{\bf 35}^\prime_0: &\quad&  \vec\lambda=(\pm 1,\pm 1,\pm 1,\pm 1)\quad
 [{\rm even}] \,,
\nonumber\\
&{}&\vec\lambda'=(\pm 1,\pm 1,0,0) \quad [+20]\,, 
\quad \vec\lambda_{1,2,3}^{\prime\prime}=(0,0,0,0) \,;\nonumber\\
{\bf 35}^{\prime\prime}_0: &\quad&  \vec\lambda=(\pm 1,\pm 1,\pm 1,\pm
1)\quad  [{\rm odd}]\,,
\nonumber\\
&{}&\vec\lambda'=(\pm 1,\pm 1,0,0) \quad [+20]\,, 
\quad \vec\lambda_{1,2,3}''=(0,0,0,0) \, ;\nonumber\\
{\bf 56}_v: &\quad & \vec\lambda=(\pm 1,\pm 1,\pm 1,0) \quad  [+24]\, ,\quad
\vec\lambda_{1,2,3}^\prime =(\pm 1,0,0,0) \quad [+6]\,; \nonumber \\
{\bf 56}_s: &\quad&  \vec\lambda=(\pm\ft32,\pm\ft12,\pm\ft12,\pm\ft12)
\quad [+3;{\rm odd}] \, ,\nonumber\\
&{}& \vec\lambda_{1,2,3}^\prime =(\pm\ft12,\pm\ft12,\pm\ft12,
\pm\ft12) \quad [{\rm even}] \,; \nonumber\\
{\bf 56}_c: &\quad&  \vec\lambda=(\pm\ft32,\pm\ft12,\pm\ft12,
\pm\ft12)
\quad  [ +3;{\rm even}]\, ,\nonumber\\
&{}& \vec\lambda_{1,2,3}^\prime =(\pm\ft12,\pm\ft12,\pm\ft12,
\pm\ft12) \quad [{\rm odd}] \,. \nonumber
\eea
%%%%%%%%%%%%%%%%%%%%%%%%%%%%%%%%%%%%%%%%%%%%%%%%%%%%%%%%%%%%%

%%%%%%%%%%%%%%%%%%%%%%%%%%%%%%%%%%%%%%%%%%%%%%%%%%%%%%%%%%%%%

\end{document}